\documentclass[fleqn,twocolumn]{revtex4-1}
\usepackage{graphicx}

\def\ga{\alpha}
\def\gb{\beta}

\def\ge{\epsilon}
\def\gg{\gamma}
\def\gd{\delta}

\def\gm{\mu}

\def\gO{\Omega}
\def\gp{\pi}
\def\gP{\Pi}

\def\gs{\sigma}

\def\gl{\lambda}

\def\dg{\dagger}
\def\delp{\partial_+}
\def\delm{\partial_-}

\def\delmu{\partial_\gm}
\def\delmuu{\partial^\gm}

\def\delrlp{\stackrel {\leftrightarrow} {\partial_+}}
\def\delrlm{\stackrel {\leftrightarrow} {\partial_-}} 
\def\delrlmu{\stackrel {\leftrightarrow} {\partial_\gm}}

\def\delrl1{\stackrel {\leftrightarrow} {\partial_1}}

\def\part{\partial}

\def\hlf{\frac{1}{2}}
\def\A0{A^{+}_0}

\def\psih{\psi_2}
\def\psihd{{\psi_2}^\dagger}
\def\psil{\psi_1}
\def\psild{{\psi_1}^\dagger}

\def\Psih{\Psi_2}
\def\Psihd{{\Psi_2}^\dagger}
\def\Psil{\Psi_1}
\def\Psild{{\Psi_1}^\dagger}
\def\Psid{{\Psi}^\dagger}

\def\Phib{\overline{\Phi}}

\def\Psib{\overline{\Psi}}

\def\psid{\psi^{\dag}}

\def\xpl{x^{+}}
\def\ypl{y^{+}}

\def\xmin{x^{-}}

\def\ulix{\underline{x}}
\def\uliy{\underline{y}}

\newcommand{\nc}{\newcommand}
\nc{\intl}{\int\limits_{-L}^{+L}\!\frac{{\rm d}x^-}{2}}
\nc{\intly}{\int\limits_{-L}^{+L}\!{{dy^-}\over\!2}}
\nc{\intlz}{\int\limits_{-L}^{+L}\!{{dz^-}\over\!2}}
\nc{\intlu}{\int\limits_{-L}^{+L}\!{{du^-}\over\!2}}
\nc{\intlv}{\int\limits_{-L}^{+L}\!{{dv^-}\over\!2}}
\nc{\intv}{\int\limits_{-V}^{}\!d^3\ulix}
\nc{\intvy}{\int\limits_{-V}^{}\!d^3\uliy}
\nc{\zmint}{\int\limits_{-L}^{+L}\!{{dx^-}\over{\!2L}}}
\nc{\zminty}{\int\limits_{-L}^{+L}\!{{dy^-}\over{\!2L}}}
\nc{\intp}{\int\limits_{0}^{+\infty}\!{{dp^+}\over\!4\gp}}
\nc{\inp}{\int\limits_{0}^{\infty}\!{{dp^+}\over{\!2\sqrt{2\gp}}}}
\nc{\inq}{\int\limits_{0}^{\infty}\!{{dq^+}\over{\!2\sqrt{2\gp}}}}
\nc{\intpp}{\int\limits_{0}^{\infty}\!{dp^+}}
\nc{\intqp}{\int\limits_{0}^{\infty}\!{dq^+}}
\nc{\intkp}{\int\limits_{0}^{\infty}\!{dk^+}}
\nc{\intlp}{\int\limits_{0}^{\infty}\!{dl^+}}
\nc{\inpp}{\int\limits_{0}^{\infty}\!{{dp^+}\over{\!2p^+\sqrt{2\gp}}}}
\nc{\inqq}{\int\limits_{0}^{\infty}\!{{dq^+}\over{\!2q^+\sqrt{2\gp}}}}
\nc{\insl}{\int\limits_{-L}^{+L}\!dx}
\nc{\intex}{\int\limits_{-\infty}^{+\infty}\!dx^1}
\nc{\inti}{\int\limits_{-\infty}^{+\infty}}
\nc{\intep}{\int\limits_{-\infty}^{+\infty}\!dp^1}
\nc{\inteq}{\int\limits_{-\infty}^{+\infty}\!dq^1}
\nc{\inteqj}{\int\limits_{-\infty}^{+\infty}\!d^1 q^1}
\nc{\intek}{\int\limits_{-\infty}^{+\infty}\!dk^1}
\nc{\intx}{\int\limits_{-\infty}^{+\infty}\!dx^-}
\nc{\inty}{\int\limits_{-\infty}^{+\infty}\!dy^-}
\nc{\intz}{\int\limits_{-\infty}^{+\infty}\!dz^-}
\def\beq{\begin{equation}}
\def\eeq{\end{equation}}
\def\bea{\begin{eqnarray}}
\def\eea{\end{eqnarray}}
\nc{\intgix}{\int\limits_{-\infty}^{+\infty}\!{{dx^-}\over\!2}}
\nc{\intgiy}{\int\limits_{-\infty}^{\infty}\!{{dy^-}\over\!2}}

\begin{document}

\title{Hamiltonian formulation of exactly solvable models and their physical  
vacuum states}
\author{L$\!\!$'ubom\'{\i}r Martinovi\u{c}}
\affiliation{BLTP JINR, 141 980 Dubna, Russia \\
and\\ 
Institute of Physics, Slovak Academy of Sciences \\
D\'ubravsk\'a cesta 9, 845 11 Bratislava, Slovakia,\\
\phantom{a}\\
{\rm and}} 
\author{Pierre Grang\'e}
\affiliation{
LPTA, Universit\'e Montpellier II, Pl. E. Bataillon\\
F-34095 Montpellier Cedex 05 France}

\begin{abstract}
We clarify a few conceptual problems of quantum field theory on the level of 
exactly solvable models with fermions. The ultimate   
goal of our study is to gain a deeper understanding of differences between  
the usual ("spacelike") and light-front forms of 
relativistic dynamics. We show that by incorporating operator solutions of 
the field equations to the canonical formalism the spacelike and light front 
Hamiltonians of the derivative-coupling model acquire an equivalent structure.  
The same is true for the massive solvable theory, the Federbush model. In the   
conventional approach, physical predictions in the two schemes disagree.     
Moreover, the derivative-coupling model is found to be almost identical  
to a free theory, in contrast to the conventional canonical treatment.  
Physical vacuum state of the Thirring model is then obtained   
by a Bogoliubov transformation as a coherent state   
quadratic in composite boson operators. To perform the same task in  
the Federbush model, we derive a massive version of Klaiber's bosonization 
and show that its light-front form is much simpler. 

\vspace{1pc}
\end{abstract}
\maketitle

The usual "spacelike" (SL) and the lightfront (LF) \cite{Dir} forms of 
relativistic quantum field  
theory (QFT) are two independent representations of the same  
physical reality. There are however striking differences between both schemes  
already at the level of basic properties \cite{LKS,LS}. This concerns the     
mathematical structure as well as some physical aspects (nature of field  
variables, division of the Poincar\'e generators into the kinematical and 
dynamical sets, status of the vacuum state, etc.)
Exactly solvable models offer an opportunity to study the structure of the 
two theoretical frameworks and their relationship since in these models 
exact operator solutions of field equations are known. From the  
solutions, essential properties like correlation functions can be  
computed nonperturbatively and independently of 
conformal QFT methods \cite{Zam}. 
Note that not all solvable models belong to the conformal class.   
Investigations of their properties in a hamiltonian approach    
permits us to study directly the role of the vacuum state 
and of the operator structures in both forms of QFT. Recall that in the LF 
scheme, Fock vacuum is often the lowest-energy eigenstate of the 
{\it full} Hamiltonian. This feature is not present in the SL theory.  

In this letter, we give a brief survey of the hamiltonian study of the 
derivative-coupling model (DCM)\cite{Schr}, the Thirring \cite{Thir} 
and the Federbush model (FM) \cite{Fed}. The unifying     
idea is to benefit from the knowledge of operator solutions of the   
field equations to re-express the corresponding Hamiltonians (both in the SL  
and LF versions) entirely in terms of true degrees of freedom which are  
the free fields. This previously overlooked aspect not only simplifies the  
overall physical picture but also removes structural differences between  
SL and LF Hamiltonians. For example, in the case of the simplest theory, the 
derivative-coupling model, the standard canonical procedure applied to the 
SL and LF Lagrangians leads to a striking result: the SL Hamiltonian contains  
an interaction term while its LF analog does not. 
Making use of exact solutions of the field equations, the SL version of 
the DCM Hamiltonian is found to have also the interaction-free form.  
Consequently, the physical SL vacuum coincides with the Fock vacuum in a   
full agreement with the LF result. However, for truly interacting models,    
the Fock vacuum is an eigenstate of the free SL Hamiltonians only.  
Interaction parts of the Hamiltonians are generally  
non-diagonal when expressed in terms of creation and annihilation 
operators. To find the true vacuum state, they have to be diagonalized.  
This is a complicated dynamical problem which however 
turns out to be tractable analytically for the Thirring and Federbush 
models. Our approach is   
to cast their Hamiltonians to the quadratic form by bosonization of the 
vector current and to    
diagonalize them by a Bogoliubov transformation, generating thereby the true  
ground state as a transformed Fock vacuum (a coherent state). We will show this  
explicitly for the Thirring model. As for the FM,
the conventional approach yields a vanishing interaction Hamiltonian  
for the LF case and a nonvanishing one for the SL case. This 
discrepancy is removed when the solution of field equations is taken into 
account leading to interaction Hamiltonians of the same structure.  
Finally, note that the solvability of the (conformally-noninvariant)  
massive FM allows one also to test the methods of CFT  
where the mass term is treated as a perturbation \cite{Zam}.      
 
{\bf Derivative-coupling model.} It is instructive to explain our main ideas  
 at a very simple theory -- a massive 
fermion field interacting with a massive scalar field via a gradient 
coupling. Its Lagrangian and field equations are  
\bea 
&&\!\!\!\!\!\!\!\!\!\!\!\!\!\!\!\!\!\!\!\!
{\cal L}=\Psib\big(\frac{i}{2}\gg^\mu\delrlmu 
- m\big)\Psi + 
\hlf \big(\delmu \phi 
\delmuu \phi  
- \mu^2 \phi^2\big) - 
g\delmu \phi J^\mu, \nonumber \\ 
&&\!\!\!\!\!\!\!\!i\gg^\mu\delmu \Psi = m\Psi + g\delmu \phi\gg^\mu \Psi, 
\\ 
&&\!\!\!\!\!\!\!\!\delmu \delmuu \phi + \mu^2 \phi = g\delmu J^\mu = 0. 
\nonumber
\label{DCML}
\eea
Here $\gg^0=\gs^1,\gg^1=i\gs^2,~\ga^1=\gg^5=\gg^0\gg^1$ and  
$\gs^i$ are the Pauli matrices. $J^\mu(x)$ is the (normal-ordered) vector 
current.  
The original Schroer's model \cite{Schr} had $\mu=0$. 
Using the notation $d\tilde{k}^1 = dk^1/\sqrt{4\gp E(k^1)}$, 
$\hat{k}.x = E(k^1)t - k^1x^1$,  
$E(k^1) = \sqrt{k_1^2 +  \mu^2}$, 
the free scalar field, quantized by $[a(k^1),a^\dagger(l^1)]=\gd(k^1-l^1)$, is  
expanded as 
\bea 
&&\phi(x) = \int\limits^{+\infty}_{-\infty}
d\tilde{k}^1\big[a(k^1)e^{-i\hat{k}.x}  
+ a^\dagger(k^1)
e^{i\hat{k}.x}\big].
\eea 
It enters into the operator solution of the equation (\ref{DCML}): 
\beq
\Psi(x) = :e^{ig\phi(x)}:\psi(x),~~i\gg^\mu\delmu \psi(x) =  m\psi(x).
\label{opsol}
\eeq
The Fock expansion of the free massive fermion field $\psi(x)$ 
\bea 
\!\!\!\!\!\!\!\!\!\!\!\!\psi(x) = \int\limits^{+\infty}_{-\infty}
{d\tilde{p}^1}\big[b(p^1)u(p^1)e^{-i\hat{p}.x}  
+d^\dagger(p^1)v(p^1)e^{i\hat{p}x}\big] 
\label{mfexp}
\eea 
contains the spinors $u^\dagger(p^1) = (\sqrt{p^-},\sqrt{p^+})$, 
$v^\dagger(p^1) = u^\dagger(p^1)\gg_5$,  
$p^\pm = E(p^1) \pm p^1$. 
The conjugate momenta
\beq
\gP_\phi = \partial_0\phi(x) - gJ^0,
~\gP_\Psi = \frac{i}{2} 
\Psi^\dagger, 
~\gP_{\Psid} = -\frac{i}{2}\Psi
\eeq
lead to the Hamiltonian $H = H_{0B} + H^{'}$, where $H_{0B}$ corresponds to 
the free massive scalar field and  
\bea  
&&\!\!\!\!\!\!\!\!\!H^{'} = \intex \big[-i\Psi^\dagger \ga^1\partial_1\Psi + 
m\Psi^\dagger
\gg^0 \Psi + g \partial_1 \phi J^1\big] \nonumber.
\eea
Since the term $(i/2)\Psib\gg^\mu\delrlmu\Psi$ in the Lagrangian is 
conventionally taken in terms of the free field,  
the first term in $H^\prime$ becomes simply $-i\psid\ga^1\partial_1
\psi$. Setting $m=0$ for simplicity, the interaction Hamiltonian acquires the 
form   
\bea  
&&\!\!\!\!\!\!\!\!\!\!\!\!\!\!\!\!
H_g = -\frac{g}{2\sqrt{\gp}}\int\limits_{-\infty}^{+\infty} 
\frac{dk^1 \vert k^1\vert^2}{\sqrt{E(k^1)\vert k^1\vert}}\Big[
a^\dagger(k^1)c(k^1) + \nonumber \\ 
&&\!\!\!\!\!\!+ c^\dagger(k^1)a(k^1) +  
a^\dagger(k^1)c^\dagger(-k^1) + a(k^1)c(-k^1)\Big],  
\label{wroh}
\eea  
where the Klaiber's bosonized representation \cite{Klaib} of the massless 
vector current was used (see Eq.(\ref{bocur}) below).  
The Hamiltonian (\ref{wroh}) is nondiagonal.  
Its diagonalization can be performed by a Bogoliubov 
transformation implemented by a unitary operator $U = \exp(iS)$ with 
\bea 
iS(\gg) = \intek \gg(k^1)\big[c^\dagger(k^1)
a^\dagger(-k^1) - H.c.\big].
\eea 
As a result, the real ground state  
has nontrivial structure, $\vert \tilde{\gO}\rangle = 
\exp\big(-iS(\gg_d)
\big)\vert 0 \rangle$ \cite{foot} 
(cf. Eqs.(\ref{cshf}-\ref{vacsim}) below). 

All this is true provided the starting Hamiltonian is the  
correct one. However, this is not the case.  
The point is that we did not use our knowledge of the operator solution 
(\ref{opsol}) in the course of the derivation of the Hamiltonian (\ref{wroh}). 
In the case of models which are not exactly solvable,  
one knows $\gg^\mu\delmu\Psi(x)$ from the corresponding Dirac equation. 
This expression should not be used in the Lagrangian since the latter would 
vanish (extremum of the action). In our case, we actually know more, namely  
$\delmu\Psi(x)$ from the solution (\ref{opsol}) and this information 
(not the free Dirac equation, however!) {\it 
should} be used in the starting Lagrangian (\ref{DCML}). This is similar 
to elimination of a nondynamical field by using its constraint in the 
Lagrangian. 
Thus, the correct procedure in our case is to insert the solution for the 
interacting field $\Psi(x)$ into the kinetic term. We find that the 
interaction cancels completely in the Lagrangian (\ref{DCML}) and we get  
$H=H_{0B}+H_{0F}$, 
\beq 
H_{0F} = \intex \big[-i\psi^\dagger\ga^1\partial_1\psi 
+ m\psi^\dagger\gg^0\psi \big].  
\eeq
Although the full Hamiltonian is simply the sum of free Hamiltonians of the  
massive scalar and fermion fields and hence the ground state of the 
DCM is just the Fock vacuum, the model is not completely  
trivial: one generates the correct Heisenberg equations 
$i\partial_0\Psi(x) = -[H,\Psi(x)] = -i\ga^1\partial_1\Psi + 
m\gg^0\Psi 
- g\partial_0\phi\Psi 
-g\ga^1\partial_1\phi\Psi$ with $H$.  
Correlation functions computed from the solution (\ref{opsol}) will 
depend on the coupling constant, too. Note also that in the usual 
treatment the momentum operator contains interaction.  
This deffect is cured in our approach.

The same picture is obtained in the LF analysis.  
Our notation is $\psi^\dg =(\psild,\psihd)$,     
$x^\mu=(\xpl,\xmin)$, $\partial_{\pm} = \partial/\partial x^{\pm}$, 
$\hat{p}.x = 1/2(\hat{p}^-\xpl + p^+
\xmin)$, $\hat{p}^-=m^2/p^+$,  
where $x^+,p^+$ and $j^\mu = (j^+,j^-)$ are the LF time, momentum 
and current. Inserting now the solution (\ref{opsol}) of the field equations  
\bea
&&2i\delp\Psi_2(x) = m\Psi_1(x) +2g\delp\phi(x)\Psi_2(x), \nonumber \\ 
&&2i\delm\Psi_1(x) = m\Psi_2(x) +2g\delm\phi(x)\Psi_1(x)
\label{lfeq}
\eea
into the LF Lagrangian 
\bea
&&\!\!\!\!\!\!\!\!\!\!\!\!\!\!\!\!\!\!{\cal{L}}_{lf}=2\delp\phi\delm\phi 
-\hlf\mu^2\phi^2 + 
i\Psihd\delrlp\Psih + i\Psild\delrlm\Psil \nonumber \\ 
&&\!\!\!\!\!\!\!\! - m\big(\Psihd\Psil +  
\Psild\Psih\big)  
-g\delp\phi j^+ - g\delm\phi j^-, 
\label{lfdcl}
\eea 
we obtain the Lagrangian of the free LF massive fermion and boson fields with 
the corresponding LF Hamiltonian 
\beq
\!\!\!\!P^- = \hlf\intx\big[m\big(\psihd\psil + \psild\psih\big) + 
\hlf\mu^2\phi^2\big]. 
\label{lfhdcm}
\eeq 
We recall that in the conventional treatment, one gets a controversial 
picture: the LF Hamiltonian still remains free while  
the SL Hamiltonian contains an interaction term  
(\ref{wroh}) and its ground state is the coherent state $\vert \tilde{\gO}
\rangle$. 

It is interesting that the analogous model with the axial vector current  
$j^\mu_{5}(x)$ is not solvable. One reason is that $j^\mu_5(x)$ is  
not conserved, and hence the (pseudo)scalar field is not 
free. More importantly, the naive generalization of the solution 
(\ref{opsol}) to $\Psi(x) = :\exp\{-ig\gg^5\phi(x)\}:\psi(x)$ actually 
does not satisfy the  
corresponding Dirac equation 
On the other hand, the Rothe-Stamatescu model \cite{RS}  
($m=0$) is exactly solvable but its Hamiltonian is actually  
a free one, contrary to the claim made in \cite{RS}.  

{\bf The Thirring model} \cite{Thir} was extensively studied in the    
past. A detailed analysis of its operator solution has been made in   
\cite{Klaib}. A systematic hamiltonian treatment based on the model's  
solvability was not given so far.  
  
The Lagrangian density of the massless Thirring model  
and the corresponding field equations read  
\bea 
&&{\cal L}=\frac{i}{2}\Psib\gg^\mu\delrlmu \Psi - \hlf gJ_\mu J^\mu,
\nonumber \\
&&i\gg^\mu\partial_\mu \Psi(x) = gJ^\mu(x)\gg_\mu\Psi(x),\nonumber \\ 
&&J^\mu=:\Psib \gg^\mu \Psi:,~~~ 
\partial_\mu J^\mu(x) = 0.
\label{TL}
\eea
The simplest classical solution is
\bea 
&&\!\!\!\!\!\Psi(x) = e^{-i(g/\sqrt{\pi})j(x)}\psi(x),
~\gg^\mu\partial_\mu\psi(x) = 0,  
\label{tsol}
\eea
implying that the interacting current $J^\mu(x)$ coincides with 
the free current $j^\mu(x)$. The latter is obtained from the operator 
$j(x)$ as  
$\sqrt{\gp}j_\mu(x) = \partial_\mu j(x)$ (see below). 

The expansion of the massless spinor field 
is the $m =0$ limit of Eq.(\ref{mfexp}). 
After the Fourier transformation, the current $j^\mu(x)=(\psid(x)
\psi(x),$$\psid(x)
\ga^1\psi(x))$ is represented  
in terms of boson operators:   
\bea 
&&\!\!\!\!\!\!\!\!\!\!\!j^\mu(x) = \frac{-i}{\sqrt{2}\gp}\inti\frac{dk^1 k^\mu}
{\sqrt{2\vert k^1 \vert}} 
\big\{c(k^1)e^{-i\hat{k}.x} - c^\dagger(k^1)e^{i\hat{k}.x}\big\}, 
\nonumber  \\ 
&&\!\!\!\!\!\!\!\!\!\!\!c(k^1) = \frac{i}{\sqrt{\vert k^1\vert}}
\inti\!\!\! dp^1\big\{
\theta\big(
p^1k^1\big)
\big[b^\dagger(p^1)b(p^1+k^1)- \nonumber \\ 
&&~- d^\dagger(p^1)d(p^1+k^1)\big]+  \nonumber \\ 
&&~+\ge(p^1) \theta\big(p^1(k^1-p^1)\big)d(k^1-p^1)b(p^1)\big\}, 
\label{bocur}
\eea
with $c, c^\dagger$ obeying the canonical  
commutation relation,  
\beq
\big[c(p^1),c^\dagger(q^1)\big] = \gd(p^1-q^1),~~c(k^1)\vert 0 \rangle = 0.
\eeq  

The Hamiltonian of the model is obtained from the Lagrangian (\ref{TL})  
after inserting the solution (\ref{tsol}) into it. The 
interaction is {\it not} canceled indicating a less trivial  
dynamics than found in the DC model. However, 
the contribution of the term  
$(i/2)\Psib\gg^\mu\delrlmu\Psi$ just reverses the sign of the 
interaction term leading to 
\beq
H = \intex \Big[-i\psi^\dagger\ga^1\partial_1 \psi - 
\frac{1}{2}g\big(j^0j^0 - 
j^1j^1\big)\Big]. 
\label{THam}
\eeq
The usual treatment gives  $+1/2g$ in Eq.(\ref{THam})    
\cite{FI1,Jap}.  
In the Fock representation, $H_0$ and $H_{g}$ have the form  
\bea 
&&\!\!\!\!\!\!\!\!\!\!\!\!H_0 = \intep \vert p^1 \vert \Big[b^\dagger(p^1)
b(p^1) + 
d^\dagger(p^1)d(p^1)\Big], \\ 
&&\!\!\!\!\!\!\!\!\!\!\!\!H_{g} = \frac{g}{\gp}\!\!\int\limits_{-\infty}^
{+\infty}\!\! 
dk^1 \vert k^1 \vert 
\Big[c^\dagger(k^1)c^\dagger(-k^1) + c(k^1)c(-k^1)\Big]. \nonumber  
\label{thmh}
\eea
$H_g$ is not diagonal and thus 
$\vert 0 \rangle$ is not an eigenstate of $H=H_0+H_g$. 
To diagonalize $H$, we form the new Hamiltonians  
$\hat{H}_0 = H_0 - T,~~\hat{H}_{g} = H_{g} + T$ \cite{MaLi},   
where $T = \intek \vert k^1 \vert c^\dagger(k^1)c(k^1)$,  
and implement a Bogoliubov transformation by the unitary operator 
$U=e^{iS}$,   
\bea
&&\!\!\!\!\!\!\!\!\!\!\!\!iS = \hlf\intep \gg(p^1)
\big[c^\dagger(p^1)c^\dagger(-p^1) - 
c(p^1)c(-p^1)\Big].    
\nonumber
\label{U}
\eea 
$\hat{H}_0$ is invariant with respect to $U$.   
$\hat{H}_{g}$ transforms non-trivially since  
$\big[S,c(k^1)\big] = i\gg(k^1)c^\dagger(-k^1)$.  
This, by $c(k^1) \rightarrow e^{iS}c(k^1)e^{-iS}$, implies 
\bea
&&\!\!\!\!\!\!\!\!\!\!\!c(k^1) \rightarrow  c(k^1)\cosh \gg(k^1) - 
c^\dagger(-k^1)\sinh \gg(k^1).  
\label{cshf}
\eea
The transformed Hamiltonian $e^{iS}\hat{H}_{g}e^{-iS}$ is diagonal,  
\beq
\!\!\!\!\!\!\!\!\!\!\hat{H}^d_{g}=\frac{1}{\cosh 2\gg_d}\intek 
\vert k^1 \vert 
c^\dagger(k^1)c(k^1),
\label{DH}
\eeq
if we set $\gg(k^1) = \gg_d = \hlf{\rm artanh}\frac{g}
{\pi}$. Thus we have  
\bea 
&&\!\!\!\!\!\!\!\!\!\!\!\!e^{iS}\big[\hat{H}_0 + \hat{H}_{g}\big]
e^{-iS}\vert 0 
\rangle = 0 
\Rightarrow 
\vert \gO \rangle = 
e^{-iS}\vert 0 \rangle.  
\label{ftv}
\eea 
The new vacuum state $\vert \gO \rangle$  
(where $\kappa =  
\hlf\tanh\gg_d$),  
\beq
\vert \gO \rangle = N\exp\Big[-\kappa\intep c^\dagger(p^1)
c^\dagger(-p^1)\Big] 
\vert 0 \rangle, 
\label{vacsim}
\eeq
corresponds to a coherent state of pairs of  
composite bosons with zero total momentum,     
$P^1 \vert \gO \rangle = 0.$
The vacuum $\vert \gO \rangle$ is invariant under  
axial $U(1)$ transformations  
\bea
&&\!\!\!\!\!\!\!\!\!\!\!\!V(\gb)\vert \gO\rangle = \vert \gO \rangle,~~
V(\gb) = e^{i\gb Q_5},  
\nonumber \\ 
&&\!\!\!\!\!\!\!\!\!\!\!\!Q_5 = \inteq \ge(q^1)\big[b^\dagger(q^1)b(q^1) - 
d^\dagger(q^1)d(q^1)\big].
\label{QQ}
\eea
Thus, no chiral symmetry breaking occurs.
This finding disagrees with the 
results \cite{FI1} where a BCS type of ansatz for the vacuum state was used.  
The true vacuum has to be  an eigenstate of the full  
Hamiltonian, however. $\vert \gO \rangle$ is such a state 
while the BCS-like state is not.  

Correlation functions can be calculated from the 
normal-ordered operator solution \cite{Klaib} 
\bea 
\Psi(x)=e^{(-ig/\sqrt{\pi})j^{(+)}(x)}\psi(x)e^{(-ig/\sqrt{\pi})
j^{(-)}(x)},  
\eea 
using the vacuum $\vert \gO\rangle$. 
$j^{(\pm)}(x)$ are positive and negative-frequency parts of the    
integrated current, $j = j^{(+)} + j^{(-)}$,  
\bea
&&\!\!\!\!\!\!\!\!\!\!j^{(-)}(x) = \frac{1}{\sqrt{2\gp}}\inteq \frac{c
(q^1)}
{\sqrt{2\vert q^1 \vert}}\theta\big( 
\vert q^1 \vert -\gl\big)e^{-i\hat{q}.x}.  
\label{jj}
\eea
$j^{(+)}(x) = j^{(-)\dagger}(x)$ and 
$\eta$ is the conventional infrared cutoff. Further 
details will be given elsewhere \cite{lupi}.   
{\bf The Federbush model} is defined by the Lagrangian 
\bea
&&\!\!\!\!\!\!{\cal L} = \frac{i}{2}\Psib\gg^\mu\delrlmu\Psi - m\Psib\Psi +   
\frac{i}{2}\Phib\gg^\mu\delrlmu\Phi - \mu\Phib\Phi - \nonumber \\ 
&&-g\ge_{\mu\nu}J^\mu H^\nu,~~\ge^{\mu\nu} = -\ge^{\nu\mu},    
\label{FEL}
\eea
which desribes two species of coupled fermion fields with masses 
$m$ and 
$\mu$ \cite{Fed}.  
Both currents $J^\mu=
\Psib\gg^\mu\Psi, H^\mu=\Phib\gg^\mu\Phi$ are 
conserved. The coupled field 
equations 
\bea 
&&i\gg^\mu\partial_\mu \Psi(x) = m\Psi(x) +g\ge_{\mu\nu}\gg^\mu H^\nu(x)\Psi(x),
\nonumber \\
&&i\gg^\mu\partial_\mu \Phi(x) = \mu\Phi(x) -g\ge_{\mu\nu}\gg^\mu J^\nu(x)
\Phi(x)
\label{FEQ}
\eea
are exactly solvable even for non-zero masses:
\bea 
&&\!\!\!\!\!\!\!\!\Psi(x) = e^{-i(g/\sqrt{\pi})h(x)}\psi(x),
~i\gg^\mu\partial_\mu\psi(x) = m\psi(x), \nonumber \\ 
&&\!\!\!\!\!\!\!\!\Phi(x) = e^{i(g/\sqrt{\pi})j(x)}\varphi(x),
~i\gg^\mu\partial_\mu\varphi(x) = \mu\varphi(x). 
\label{Fsolu}
\eea
In quantum theory, the  above exponentials  
have to be regularized by the "triple-dot ordering" \cite{Wight}. 
The potentials $j(x), h(x)$ given in terms of the free 
currents as $\delmu j(x) = \sqrt{\gp}
\ge_{\mu\nu}j^\nu(x)$, 
$\delmu h(x) = \sqrt{\gp}\ge_{\mu\nu}h^\nu(x)$ enter into 
the solutions (\ref{Fsolu}) in an "off-diagonal" way.  
After inserting the solutions into the Lagrangian 
(\ref{FEL}), the interaction term changes its sign leading to the 
Hamiltonian 
\bea 
H = H_0 +g\intex
\big(j^0h^1 - j^1h^0\big).
\label{FEH}
\eea 
where $H_0$ is the sum of two free fermion Hamiltonians.

The LF field equations are also solved by (\ref{Fsolu}) with 
the free LF fields $\psi(x)$, $\varphi(x)$; $j(x),h(x)$  
are given by 
$2\delm j(x) = \sqrt{\gp}j^+(x)$, 
$2\delm h(x) = \hlf\sqrt{\gp}h^+(x)$. 
In the standard LF  
treatment, one would simply insert the solution of the 
fermionic constraint into ${\cal L}$. This yields however 
the free LF Hamiltonian! It is only after inserting the full  
solution like in the SL case that one obtains  
the four-fermion interaction term also in the LF case:
\beq
P^-_g=\hlf g\intx \big(j^+ h^- - j^-h^+\big).
\label{Flfh}
\eeq   
The interacting SL Hamitonian (\ref{FEH}) contains terms 
composed solely from creation  
or annihilation operators, so the Fock vacuum is not its 
eigenstate. The diagonalization can be performed by a Bogoliubov  
transformation using a {\it massive} current 
bosonization. This is considerably more complicated than 
the massless case \cite{Klaib} but the generalized  
operators $c(k^1)$ can be derived \cite{lupi}.  
The analogous LF operators $A,A^\dagger$ are much simpler and have a 
structure similar to the massless SL case (\ref{bocur}). The bosonized 
LF current is similar to the SL current (\ref{bocur}):    
\bea
\!\!\!\!\!\!\!\!\!\!j^{+}(x) = 
\frac{1}{4\gp}\int\limits_0^\infty \frac{dk^+}{\sqrt{k^+}}~k^+ 
A(k^+,\xpl)e^{-\frac{i}{2}k^+\xmin} + H.c.    
\label{lfmasb} 
\eea
Due to $[A(k^+),A^\dagger(l^+)] = \gd(k^+ - l^+)$, valid at 
$\xpl=\ypl$, the solution  
(\ref{Fsolu}) can be easily normal-ordered:  
\bea 
&&\!\!\!\!\!\!\!\!\!\!\!\!\!\!\Phi(x) = \exp\big\{-i\frac{g}
{\sqrt{\gp}}
\hat{A}^\dagger(x)
\big\}
\exp\big\{-i\frac{g}{\sqrt{\gp}}\hat{A}(x)\big\}\varphi(x), 
\nonumber \\ 
&&\!\!\!\!\!\!\!\!\!\!\!\!\hat{A}(x) = \frac{1}{4\gp}
\int\limits_0^\infty
\frac{dk^+}{\sqrt{k^+}}
A(k^+,\xpl)e^{\frac{i}{2}k^+\xmin}.    
\label{lfnos}
\eea
Similar formulae hold for 
the solution $\Psi(x)$ built from the operators  
$B(k^+,\xpl), B^\dagger(k^+,\xpl)$ which are constructed from 
$h^+(x)$.
The $j^-$ and $h^-$ currents contain the boson 
operators $C(k^+,\xpl), D(k^+,\xpl)$ 
and their conjugates, related to 
$A,A^\dg,B,B^\dg$ via the current conservation. In contrast to its SL 
analog, the interacting LF Hamiltonian is diagonal and therefore 
$ \vert 0 \rangle$ is  
its lowest-energy eigenstate:   
\bea 
&&\!\!\!\!\!\!\!\!\!\!\!\!\!\!\!\!P^-_g = \frac{g}{8\gp}\intkp k^+ 
\Big[A^\dg(k^+)D(k^+) + D^\dg(k^+)A(k^+)   
\nonumber \\   
&&~~~~~~~~~ - B^\dg(k^+)C(k^+) 
-C^\dg(k^+)B(k^+)\Big].
\label{lfhf}
\eea 
Diagonalization of the bosonized SL Hamiltonian 
yielding the true SL vacuum state $\vert \gO \rangle$ 
will be given in \cite{lupi}.  

The next step will be to compute the correlation functions in 
both schemes. This task is not simple since one needs 
to know the commutators of the composite boson operators at 
unequal times \cite{lupi}. This is the place where complexities 
of the usual triple-dot ordering technique \cite{ST} enter into our 
bosonization approach. 
The knowledge of exact  
correlation functions will allow us to get a deeper insight into 
the relation between the SL and LF forms of the 
Federbush model and of QFT in general.


{\bf Acknowledgements.} This work was supported by the VEGA grant 
No.2/0070/2009, by the Slovak CERN Commission and by IN2P3 funding at the 
Universit\'e Montpellier II.  


\begin{thebibliography}{99}
\bibitem{foot}Our discussion here is a little heuristic for 
simplicity. Some operators have to be regularized by a cutoff  
to exist mathematically. A more rigorous treatment with the same physical 
conclusions based on fields as operator-valued distributions \cite{GW} 
will be given in \cite{lupi}. Also, our results remain unaltered when the 
products of  
the fermion fields are regularized by point-splitting.
\end{thebibliography}
\end{document}